\documentclass[a4paper]{article}
\usepackage{graphicx}
\begin{document}

\title{Anomaly-driven decay of massive vector bosons}
\author{Alejandro Rivero\thanks{Zaragoza University at Teruel.  
           {\tt arivero@unizar.es}}}

\maketitle

\begin{abstract}

We inquire if the total decay rate of neutral vector mesons can contain
 relevant off-shell contributions
from the  triangle anomaly.

The answer seems to be affirmative for the case of J/$\Psi$, and
as a byproduct we get estimates of the decay width for neutral pions.
\end{abstract}
 
A vector meson can not decay into two real photons. But that doesn't mean that
the decay do not exist, it inhabitates the virtual channels. When we rescale the
pion anomalous decay into other pseudoscalars, we get an estimate of the 
$\gamma \gamma$ decay of these particles. But when we scale up to vector mesons,
we can not get real decay. Instead, the virtual decay will be a contribution
to the total decay amplitude.
We wonder in what cases, if any, this contribution will be the dominant one. 

We can
ask the question in a reverse way: to scale down, so that a dominant role of the
triangle in the massive vector will amount to an estimate of the anomalous 
decay of $\pi^0$ (and, from $\Gamma_{\pi^0}$, also of the constant $f_{\pi^0}$)

Lets explore the experimental data. From PDG 2004

$$
\Gamma_{\pi^0}^{exp} = 7.8 \mbox{ eV } (+0.6,-0.5,S=3)
$$

or, from second order perturbation theory \cite{pert},
 
$$
\Gamma_{\pi^0}^{th} = 8.10 \, eV \,(\pm0.08)
$$

The $\rho,K^{*},\omega,\phi$ have access to stronger decay mechanisms via OZI-allowed transitions, so the contribution from the triangle anomaly goes unnoticeable and we can not make use of them \footnote{Still, we could make some use of decays of $\rho$ to photon plus anything, or of other mesons to photon plus charged plus anything, but a very subtle analisis of the decay process is needed}. Only for massive quarkonia, $J/\Psi$ and $\Upsilon$, the transitions
allowed by OZI rule are closed by the respective energy thresholds of $D\bar D$ and $B\bar B$.

So our first candidate to try is $J/\Psi$. We get
$$
 ({m_{\pi^0} \over m_{J/\Psi}})^3  \; \Gamma_{J/\Psi}  =  7.5 \; eV \;(\pm 0.3) 
$$
which is a fairly perfect quantity, practically exhausting the available width.

On the contrary for $\Upsilon$, the decay rate is too low. 
Perhaps something is closing the access
to the triangle. The decay fails by about one MacGregor's magnitude\cite{mcG}\footnote{
Experimentally, three jets are observed. Does it decay via three gluons, as expected
from OZI violation, or via some box diagram? Or direct quark decay? (the order of magnitude
seems right for the later possibility)}\footnote{Note also that the equidistance with the electroweak decaying pseudoscalars is maintained approximately}.
It can be a hint of the special role of the third generation; for the top quark it is
unnatural to assume a zero mass, and via SU(2) -or isospin if you prefer- the protection
against anomaly extends to the bottom quark.

And now, a
surprise! For $Z^0$, we can claim that
a 't Hooft principle seems to be working behind the curtains, because we get an 
almost excessively exact

$$
 ({m_{\pi^0} \over m_{Z^0}})^3  \; \Gamma_{Z^0}  =   8.1 \; eV \;(\pm 0.0)
$$

From this last result, it should mean that we can get $f_{\pi^0}$ from only the electroweak
parameters (plus $m_\pi$ in higher orders). Lets write it explicitly, for instance
by joining formulae (1.3) and (8.91) of \cite{Ellis:1991qj}
$$
\xi^2 ({\alpha\over\pi})^2 {1 \over 64 \pi }{1 \over f_\pi^2} 
\approx {G_F \over 6 \sqrt 2 \pi} \sum_f C_f (|V_f|^2 + |A_f|^2)
$$

Jointly, the three results could be interpreted as hints of a deeper symmetry that 
affects to the Z particle too and at the same time protects the top/bottom quarks from
getting a zero mass. About how this zero mass could appear, let me to address you
towards the figure in \cite{rivero} where it was suggested to drive the fine structure 
constant down to zero at the same time that the masses of the  first generation\footnote{In
an intermediate step of the process, the mass of the muon becomes degenerate with the 
pion, an amusing "reflected susy". The pion becomes stable, and then -if having
isospin symmetry- it still can get a non null mass because of the infinite in $1/f_{\pi^+}$}.
 We also suspect that the second
generation should get a null mass against the electroweak scale, from hints in \cite{hans}.

In a humbler approach, let me note that the discrepancy between theory and calculation
for hadronic OZI-forbidden rates has been traditionally problematic, see for instance 
section 4.5.3 of \cite{LeYaouanc:1988fx}; thus the possibility of a intermediate virtual
state could help to settle the issue.

{\bf About the figures:} This work was inspired by the classification of decays reviewed
in \cite{mcG}, where integer powers of the fine structure constant are used, without accounting
the mass nor another kinematic. Most of this impressive ordering gets lost when we
scale with a power of the mass, so our graphs are not so interesting, and they only show the
peculiarity of the $\Upsilon$ particles below the $B\bar B$ threshold. To put some separation
in the horizontal axis, I have preferred the Inverse ArcTan visualisation communicated
to me by Yuri Danoyan some weeks ago, but taking as mass origin the $\eta'$ instead of the
nucleon mass. Yuri calls this parametrisation "dodecahedron" because it produces some
angles about $180-144=36$ and $180-108=72$ degrees, as well as submultiples.

\begin{figure}
 \centering
 \includegraphics[width=11cm]{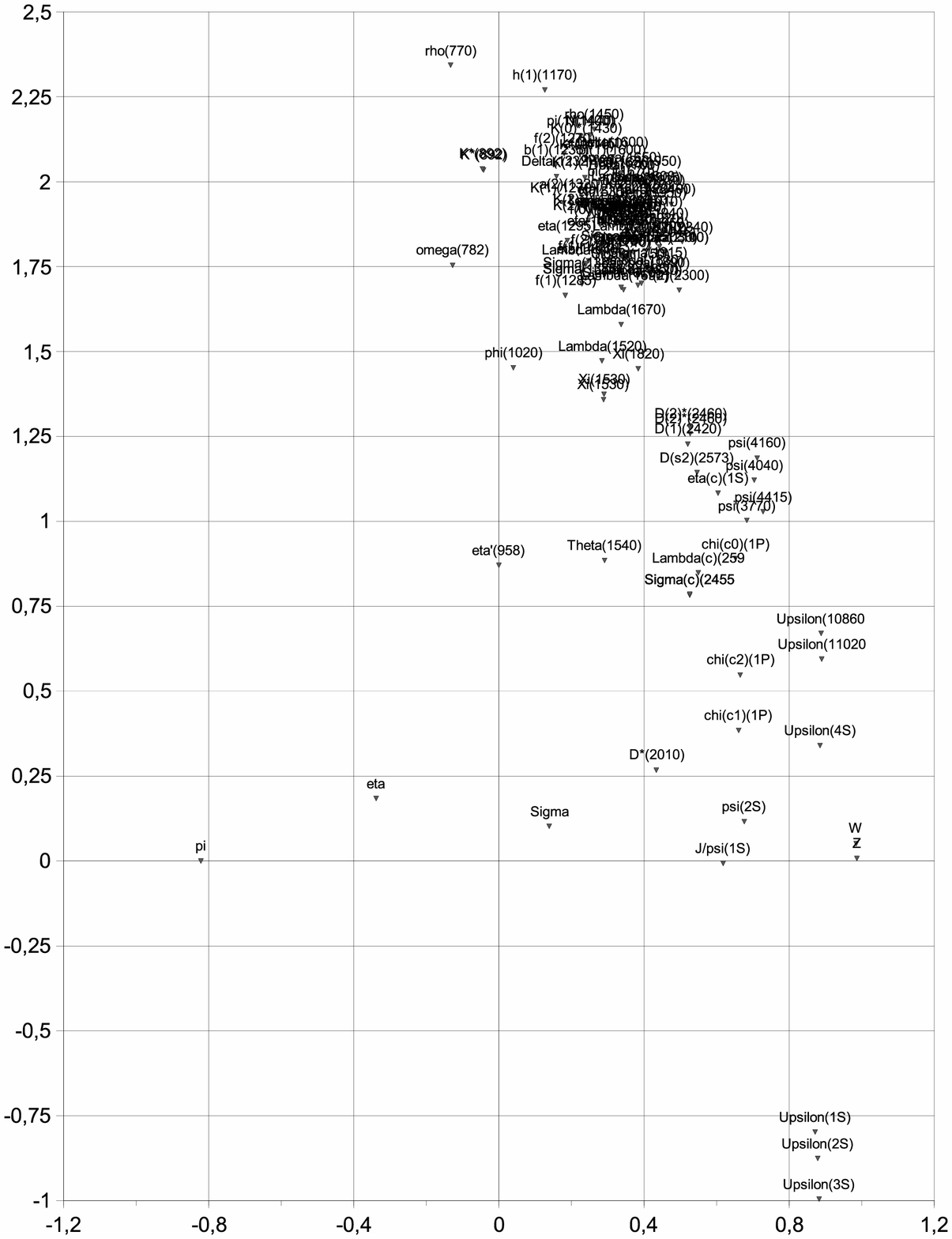}
 \caption{Decay widths of PDG 2004 table of particles, scaled according the cube of their
 mass, as described in the text. Electroweak particles are shown in the next figure, for
 completeness. We use Danoyan's parametrisation in the horizontal axis, and
  MacGregor's in the vertical}
\end{figure}

\begin{figure}
 \centering
 \includegraphics[width=11cm]{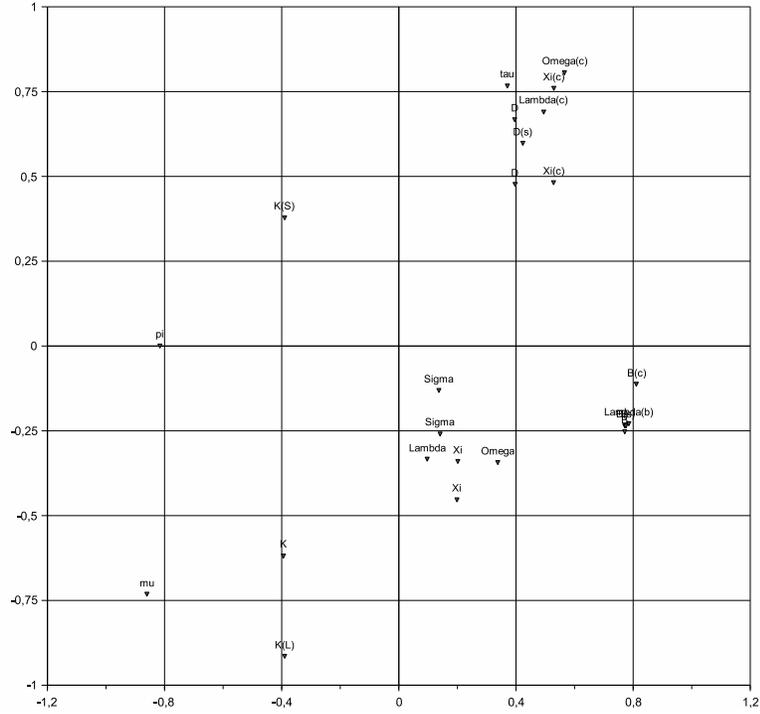}
 \caption{Decay widths for the electroweak decaying particles, scaled according $M^3$ as in
 previous plot. Note that the pion in this plot is the charged one}
\end{figure}

\end{document}